\documentclass{article}
\usepackage{bifchaos}
\usepackage{multicol}
\usepackage{vmargin}
\usepackage{latexsym,float,flafter,balanced,subfigure,longtable}
\usepackage{epsfig}

\setpapersize{A4}
\setmarginsrb{1.3cm}{2cm}{2cm}{2.5cm}{0pt}{0mm}{0pt}{0mm}

\newcommand{\BE}{\begin{equation}}
\newcommand{\ee}{\end{equation}}
\newcommand{\BC}{\begin{center}}
\newcommand{\EC}{\end{center}}
\newcommand{\BI}{\begin{itemize}}
\newcommand{\EI}{\end{itemize}}
\newcommand{\BA}{\begin{eqnarray}}
\newcommand{\EA}{\end{eqnarray}}

\begin{document} 

\title{\bf
Spatiotemporal Chaos, Localized Structures and Synchronization in 
the Vector 
Complex Ginzburg-Landau Equation
}

\author{
Emilio Hern\'andez-Garc{\'i}a$^1$, Miguel Hoyuelos$^1$, \\ 
Pere Colet$^1$, Ra\'ul Montagne$^2$\thanks{Present Address: SCRI, FSU, 
Tallahassee (Florida, USA)}, and Maxi San Miguel$^1$ \\ 
$^1$Instituto Mediterr\'aneo de Estudios Avanzados 
IMEDEA\thanks{URL: {\tt http://www.imedea.uib.es/Nonlinear}} \
(CSIC-UIB) \\
Campus Universitat de les Illes Balears, E-07071 Palma de Mallorca (Spain) \\
$^2$Instituto de F\'\i sica, Facultad de Ciencias\\
Igua 4225 C.P. 11400, Montevideo (Uruguay)}
 
\date{July 30, 1998}

\maketitle

\begin{abstract}  
We study the spatiotemporal dynamics, in one and two spatial dimensions, of two 
complex fields which are the two components of a vector field satisfying a 
vector form of the complex Ginzburg-Landau equation. 
We find synchronization and generalized synchronization of the spatiotemporally
chaotic dynamics. The two kinds of synchronization can coexist simultaneously 
in different regions of the space, and they are mediated by localized 
structures. A quantitative characterization of 
the degree of synchronization is given in terms mutual information 
measures.

{\tt To appear in International Journal of Bifurcation and Chaos (1999). 
A version
with higher quality figures could be found at 
http://www.imedea.uib.es/PhysDept/publicationsDB/date.html}

\end{abstract} 



\begin{twocolumns}

\section{Introduction} 
\label{intro}

Starting from the pioneering experimental study by Huygens with two marine
pendulum clocks hanging from a common support \cite{huygens}, synchronization
phenomena have been subject of intense study in many physical  and biological
systems \cite{winfree,strogatz}.  Since \cite{pecora90} established that chaotic
oscillators can also become synchronized,  many applications and extensions of
the original idea have been identified. Some of them are the  possibilities of
partial (i.e. {\sl phase}) synchronization \cite{rosenblum}, generalized 
synchronization
\cite{rulkov,kocarev}, and synchronization of spatiotemporally chaotic systems
\cite{prl}. A natural class of systems in which synchronization of
spatiotemporal chaos can be  explored is the one constructed in the following
way \cite{kocarevstc}: Take a couple of chaotic  systems that synchronize when 
appropriately
coupled. Then make copies of this  composite system, one copy at each point of 
a spatial lattice, and couple them spatially to study how the spatial coupling
modifies the synchronization characteristics. This class of systems displays a
kind of spatiotemporal chaos, and of chaos synchronization,  which is a natural
extension of the one obtained from the single  composite system.  

But there are spatially extended systems that display a kind of spatiotemporal
chaos {\sl produced} by the spatial coupling, so that chaos is absent from them
in spatially homogeneous situations.  Spatiotemporal chaos has  thus a rather
different origin which could lead eventually to a different  kind of
synchronization. The Complex Ginzburg-Landau (CGL) equation is one of  such
model systems: in the absence of spatial coupling, it is simply the normal 
form for a Hopf bifurcation, so that chaotic behavior is absent. But spatial 
coupling induces a huge variety of intricate chaotic behavior 
\cite{Shraiman,Chate94n,Chate94d,Eguiluz}. The possibility of synchronized
chaos between a pair of amplitudes satisfying a pair of coupled CGL equations 
in one spatial dimension was considered in
\cite{prl}. A kind of generalized synchronization was found and characterized.
In this Paper we show that different kinds of synchronization are possible, all
mediated by the presence of localized objects in the equations solutions, which
become specially robust in twodimensional situations. For small couplings, 
both usual and generalized synchronization coexist simultaneously in different
regions of the space. For larger couplings only generalized synchronization 
remains. 

\section{Vector Complex Ginzburg-Landau equation in the one dimensional case}
\label{1dsection}

In the context of amplitude equations, the CGL  is the generic model describing
slow modulations in the oscillations of spatially coupled oscillators close to
a Hopf bifurcation \cite{vSaarloos}. Pairs of coupled CGLs have been derived in
a variety of contexts, mainly related to the interaction of counterpropagating
waves  \cite{ToniDanielPRL}. A different context in which coupled CGLs appear
is in optics: the interaction of the two polarization states of light in large
aperture lasers  has been show to be described by coupled CGLs with particular
symmetries which  allow the pair to be thought as a Vector Complex
Ginzburg-Landau (VCGL)  equation \cite{maxiPRL}.  The two components of the
VCGL equation can be written as 
\BA
\partial_t A_+ &=& A_+ 
+(1+i\alpha) \nabla^2 A_+ \nonumber \\ 
&&- (1+i\beta)\left(|A_+|^2 + \gamma|A_-|^2 \right) A_+ \nonumber \\
\partial_t A_- &=& A_- 
+(1+i\alpha) \nabla^2 A_- \nonumber \\ 
&&- (1+i\beta)\left(|A_-|^2 + \gamma|A_+|^2 \right) A_- \, .
\label{vcgl}
\EA
$A_+$ and $A_-$ are the two components of the vector complex field, which in
optical applications are identified with the left and right circularly 
polarized 
components of light. $\alpha$ and $\beta$ are real parameters. We will
restrict our discussion to the case in which $\gamma<1$ is a real number and
$1+\alpha \beta>1$ is
satisfied (Benjamin-Feir  stable range). These restrictions appear for laser 
systems preferring linearly polarized emission \cite{maxiPRL}. 
Note that, different from other contexts  in which coupled CGL 
Eqs. have been derived, in the optical polarization context a group-velocity
term is absent from (\ref{vcgl}). An extensive analysis of coupled CGL 
equations for  
other parameter regimes can be found in \cite{vHvS}. 

Within the Benjamin-Feir stable range, there is a region of parameters for
which  a single CGL Eq. has still a regime  of spatiotemporally chaotic
behavior. It is the so called {\sl spatiotemporal intermittency} regime
\cite{Chate94n}. States in this regime consist in patches of travelling waves
interrupted by localized objects (depressions or holes in the field amplitude)
that move rather erratically around the system while emitting waves and
perturbations. These  {\sl homoclinic holes} \cite{vH} are the responsible of
sustained spatiotemporal chaos in the system. Both nonlinear dispersion (the
$\beta$ parameter) and spatial coupling are needed to obtain this kind of
chaotic behaviour. Figure \ref{gamma0} shows a spatiotemporal  configuration of
the modulus of the field, obtained from one of the two equations in
(\ref{vcgl}) when $\gamma=0$, so that it is uncoupled to the other component.
Blue lines are the trajectories of  the localized depressions   disorganizing
the system, whereas green-yellow regions are laminar states. When two coupled
equations with $\gamma \ne 0$, are considered, new objects come into play. 
Fig. \ref{pareja1d} shows the modulus of the two components  $|A_+|$ and
$|A_-|$ for  $\gamma=0.7$. In addition to the blue holes there also red maxima
in the amplitudes. It is clear that the maxima in one of the components appear 
where the other component has a hole, so that a strong anticorrelation is 
present between these variables.  

\begin{figure}[H]
\begin{center}
\epsfig{file=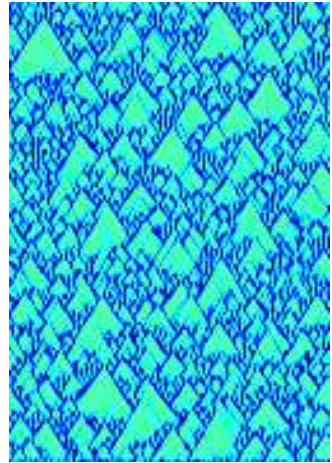,width=.24\textwidth}
\caption[1]{Spatiotemporal evolution of the modulus squared $|A(x,t)|^2$ of the 
scalar CGL equation ($\gamma = 0$) for $\alpha=0.2$ and $\beta=-2$.  Time is 
running upwards from $t=0$ to $t=400$ and $x$ is in the horizontal direction.
The system size is $L=512$.}
\label{gamma0}
\end{center}
\end{figure}

\begin{figure}[H]
\begin{center}
\epsfig{file=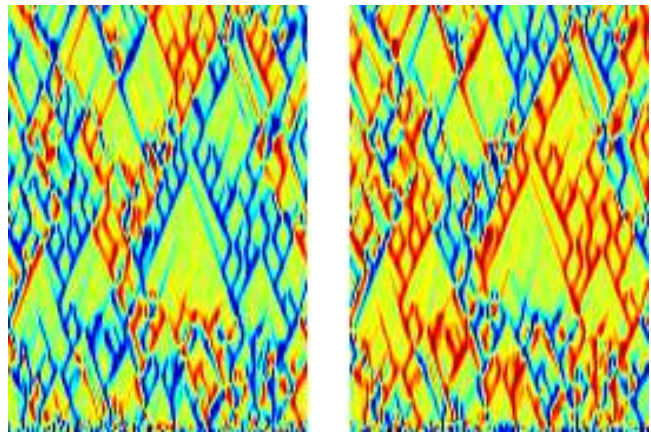,width=.48\textwidth}
\caption[2]{Spatiotemporal evolution of $|A_+(x,t)|^2$ (left) and
$|A_-(x,t)|^2$ (right) for $\gamma = 0.7$. The other parameters are as in Fig.
\ref{gamma0}}
\label{pareja1d}
\end{center}
\end{figure}

When analyzed separately, both $|A_+|$ and $|A_-|$ display spatiotemporal
chaos. The knowledge of one of these variables, however, gives a large amount
of information on the other, since they are strongly anticorrelated. This is 
precisely the content of the concept of {\sl generalized synchronization}
\cite{kocarev,rulkov}: the two chaotically evolving variables are not identical
as in the usual synchronization case, but there is a functional relationship
between them which allows close prediction of  one of them when the other is
known. In \cite{prl} the functional relation was  identified and a mutual
information calculation showed that the generalized synchronization became more
perfect as the parameter $\gamma$ approached $\gamma=1$ from below. For
$\gamma=1$ the functional relation between the  two fields is $|A_+|^2 +
|A_-|^2 = 1$. The appearance of correlations between the components is clearly
mediated by the presence of the localized objects, so that this is a kind of
generalized synchronization of chaos specific of spatiotemporal chaos, 
that is, it is
absent in systems without spatial dependence. 

It should be noticed that only the moduli $|A_+|$ and $|A_-|$ develop
correlations and we thus find amplitude synchronization. The corresponding 
phases do not become synchronized. This 
is the case opposite to the one commonly observed of {\sl phase synchronization}
\cite{rosenblum}. The reason for this is that the coupling 
between the fields in Eq. \ref{vcgl} involve only the moduli.

\begin{figure}[H]
\begin{center}
\epsfig{file=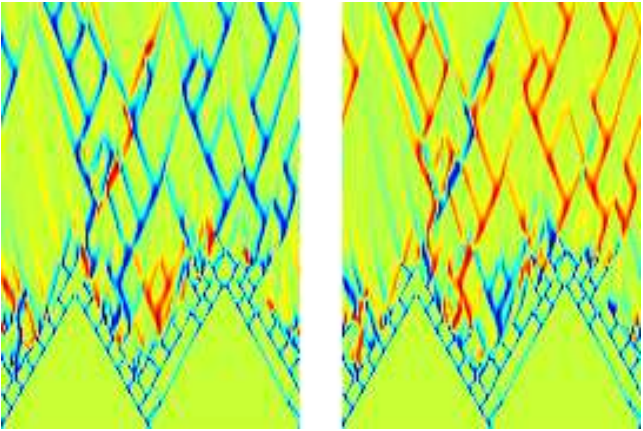,width=.48\textwidth}
\caption[3]{Temporal evolution of $|A_+(x,t)|^2$ (left) and $|A_-(x,t)|^2$
(right) starting from an initial condition where 
$A_+(x,0) \approx A_-(x,0)$. $\alpha
= 0.6$, $\beta = -1.4$ and $\gamma = 0.7$.  The other parameters are as in Fig.
\ref{gamma0}}
\label{new1d}
\end{center}
\end{figure}

Although the meaning and origin of the generalized synchronization found is 
clear, one often prefers to apply the term ``synchronization of chaos'' 
to situations closer to the original ideas \cite{pecora90}, which in the 
present context
would mean a tendency of the two fields involved to take identical values, not
just anticorrelated ones. We have explored, numerically and analytically, 
this possibility and found that 
there exist states in which the amplitude of the two moduli are identical. 
They are however dynamically unstable, so that the system does not approach 
them unless particular initial conditions are carefully selected. One example 
is shown in Fig. \ref{new1d}: during the early part of the evolution the two
components are perfectly synchronized, but a decay to the previous
anticorrelated case occurs at later times. Further states will be discussed
elsewhere. Two types of localized structures can be seen in Fig. \ref{new1d},
the ones for which both fields have simultaneously a hole 
and the ones in which a hole in one of the fields is associated to a pulse in the
other field. Hole-hole and hole-pulse localized structures will also be present
in twodimensional systems as we will show in the next section. Furthermore,
we will see that topological restrictions can make 
stable in higher dimensions the hole-hole localized objects related to the ones
appearing at the early times in Fig. \ref{new1d}.

\section{Twodimensional defects}
\label{2dsection}

The onedimensional localized structures of the previous section become
topological defects in two dimensions. They have been introduced in \cite{Gil}
and properly classified in \cite{Pismen92,Pismen94}. There are  two main types
of topological defects in (\ref{vcgl}): vectorial defects are  objects for
which the two amplitudes become identical near the defect core, where both
vanish. They are the two dimensional analogs of the hole-hole structures
present in the early part of Fig. {\ref{new1d}. In the other class of 
defects only one 
of the  amplitudes vanishes, whereas the other presents a maximum. These are
the twodimensional analogues of the hole-pulse localized structures in Fig.
\ref{pareja1d} and  in Fig. \ref{new1d} for larger times. Fig. \ref{2d}(a)
shows the amplitude of one of the two fields containing both types of defects.
The vectorial ones emit waves that entrain a whole domain around them whereas 
defects in which 
only one amplitudes vanishes behave more passively and remain at the domain
borders. In fig. \ref{2d}(b) the global phase $\phi_g = \phi_+ + \phi_-$, where
$\phi_+$ and $\phi_-$ are the phases of $A_+$ and $A_-$, is plotted.  Clearly, 
there are 
two different kinds of vectorial defects. One is produced by two defects of the
same topological charge, which correspond to the two-armed spiral in the plot
of $\phi_g$. The other is generated with defects of opposite charges, leading
to a target pattern in the plot of $\phi_g$.

In the dynamical evolution from random initial conditions, vectorial defects
are spontaneously formed  for the set of parameters of Fig. \ref{2d}. This is
the same set of parameters explored in \cite{prl} for a  1D problem and it
leads here to a glassy or frozen configuration like the one shown. However, 
increasing the value of the coupling parameter $\gamma$, the vectorial defects
become unstable. The  system evolves then in a disordered dynamics which is
governed by defects in which only one  amplitudes vanishes. The number of
defects is conserved, after a transient regime, during very long  times of
evolution. A snapshot of this state is shown in Fig \ref{mixdyn},
where it is seen that  a zero of one of the amplitudes corresponds to a maximum
of the other one. These defects, strongly  anticorrelated in amplitude, move 
together in time so that the kind of spatiotemporal chaos they sustain 
displays generalized synchronization between the two amplitudes
$|A_\pm|$.

\begin{figure}[H]
\begin{center}
\epsfig{file=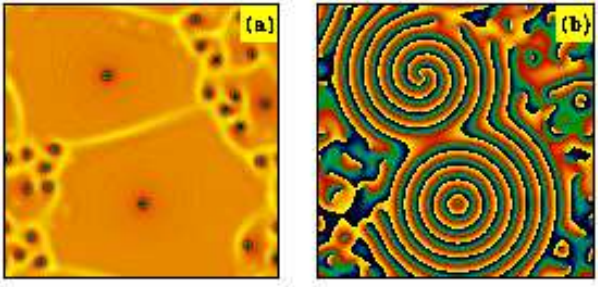,width=.5\textwidth}
\caption{
Field configuration showing different kinds of defects in two spacial
dimensions.  One of the amplitudes, $|A_+|^2$ (a), and the global phase $\phi_g$
(b), at a given time are shown. Parameters used: $\gamma =0.1$, 
$\alpha = 0.2$ and $\beta = 2$, system size 128$\times$128.}
\label{2d}
\end{center}
\end{figure}

\begin{figure}[H]
\begin{center}
\epsfig{file=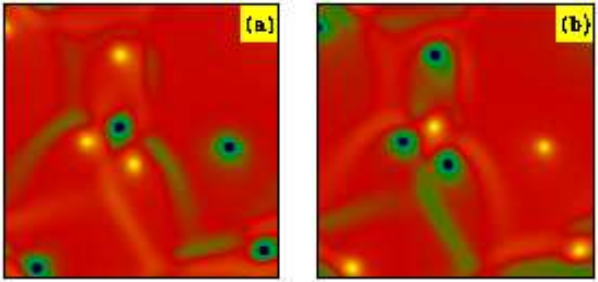,width=.5\textwidth}
\caption{Snapshot of the intensities of the two fields (a) $|A_+|^2$ and (b)
$|A_-|^2$ for $\gamma=0.8$  and $\alpha$ and $\beta$ as in Fig \ref{2d}. A dark
blue (yellow) dot corresponds to a zero (maximum) of  the amplitude.}
\label{mixdyn}
\end{center}
\end{figure}

A quantitative description of the process of increasing generalized 
synchronization 
as $\gamma$ is  increased can be given in terms of information measures as
already introduced for the 1D case in  \cite{prl}. This description also
identifies the transition from glassy to dynamic states due to the  instability
of the vectorial defects. The joint probability density $p(|A_+|,|A_-|)$ is
plotted in fig. \ref{sincr} for different values of $\gamma$, as a 3D plot and,
in the same figure, as a density plot. The density plot is obtained taking the
simultaneous values of $|A_+|$ and $|A_+|$ at  different space-time points. For
a small coupling, $\gamma = 0.1$, we obtain a diffuse cloud of points with a
broad maximum around  $|A_+| \simeq |A_-| \simeq 1$ with deviations from these
values being uncorrelated, except for the  points laying in the line $|A_+| =
|A_-|$. The presence of such line in plots of coupled variables is the 
classical 
signature of conventional synchronization. The points on the line correspond 
to the core of the vectorial defects in  which 
the two amplitudes take the same value. For $\gamma \simeq 0.3$, the cloud of
points broadens and the line $|A_+| = |A_-|$ becomes diffuse  and disappears.
This behavior identifies the instability of the vectorial defects. Such
instability appears in a narrow range of $\gamma$ and with different mechanisms
for the two types of vectorial defects.  As the coupling is further increased
($\gamma=0.5$ and 0.95), the cloud of points approaches the  curve given by
$|A_+|^2 + |A_-|^2 = 1$ as in one dimension. This indicates anticorrelation
between $A_+$ and $A_-$ and a generalized amplitude synchronization of the 
dynamics 
which increases  with $\gamma$. Similar qualitative behavior with $\gamma$ is
observed for other values of $\alpha$  and $\beta$ for which a glassy state
occurs at $\gamma=0$.

\begin{figure}[H]
\begin{center}
\psfig{figure=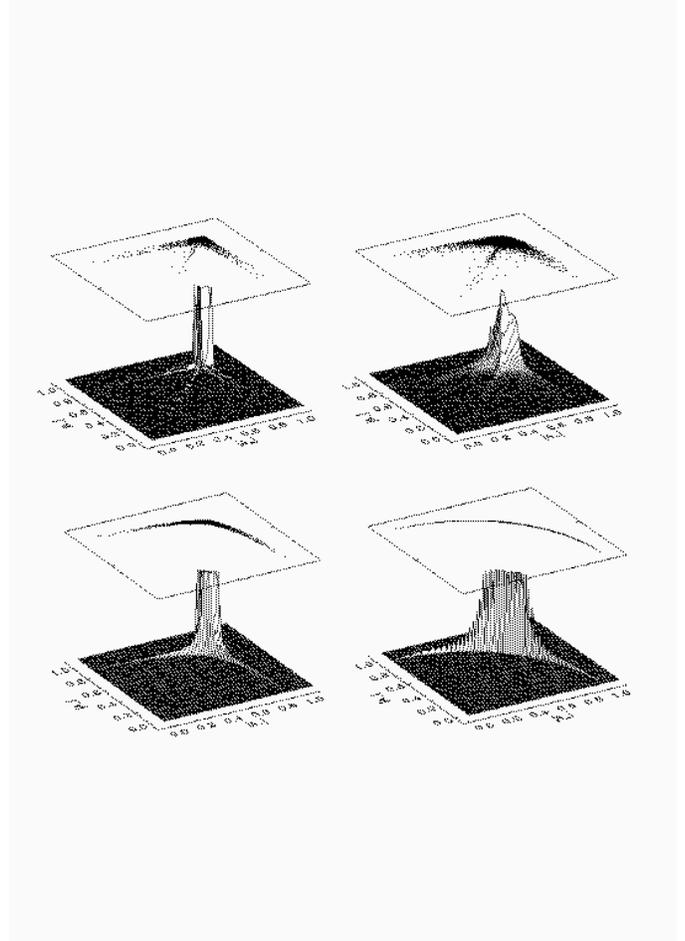,width=.5\textwidth}
\caption{Joint probability distribution $p(|A_+|,|A_-|)$, for $\alpha$ and
$\beta$ as in Fig \ref{2d},  shown as a 3D surface for different values of
$\gamma$.  From top to bottom and from left to right, $\gamma = 0.1$, 0.3, 0.5
and 0.95.  On top of each surface the corresponding plots of $|A_+(x,t)|$ vs
$|A_-(x,t)|$ are shown (the density of dots is proportional to
$p(|A_+|,|A_-|)$). A time average was taken over 100 samples separated $\Delta
t = 1$. The 3D surfaces for $p(|A_+|,|A_-|)$ has been cut at 
$p=0.0025$.}
\label{sincr}
\end{center}
\end{figure}

Two quantities that can be extracted from the probability density are the
entropy for a single amplitude  and their mutual information. The entropy is
defined as $H(X) = - \sum_x p(x) \ln p(x)$, where $p(x)$ is the probability
that $X$ takes the value $x$. $H(X)$ measures the randomness of a discrete
random variable $X$. For two random discrete variables $X$ and $Y$, with joint
probability density $p(x,y)$, the mutual information $I(X,Y) = -\sum_{x,y}
p(x,y) \ln [p(x) p(y)/p(x,y) ]$ gives a measure of the statistical dependence
between both variables; the mutual information being 0 if and only if $X$ and
$Y$ are independent. Considering the discretized values of $|A_+|$ and $|A_-|$
at space-time points as random variables $X=|A_+|$, $Y=|A_-|$, their mutual
information is a measure of their generalized synchronization.  The dependence 
of the 
entropy of $A_+|$ and $|A_-|$ and their mutual information $I$ on the coupling
parameter $\gamma$, shown in Fig. \ref{mutua} identifies three  regimes: For
low values of $\gamma$ there is a glassy or frozen configuration dominated by
large  islands around vectorial defects. This relatively ordered state gives a
relatively small value of $H$.  The mutual information is not large because of
the weak coupling between the fields. A second regime is
associated  with the instability of the vectorial defects. This is highlighted
by a maximum value of the entropies  and a minimum of $I$ for $\gamma \simeq
0.3$. This instability first disorders the configuration by  reducing the size
of the domains. This yields an increase of the entropies, but this disorder is
not  correlated in both components as indicated by the decrease of $I$. For
higher values of $\gamma$ ($\gamma > 0.35$) we enter the third regime
characterized by disordered  configurations which evolve in time like the one
shown in Fig. \ref{mixdyn}. The disorder of each  amplitude $|A_\pm|$ is
measured by a relatively large value of $H$, while the increasingly 
synchronized dynamics is measured by an increasing value of $I$ which
approaches its maximum  possible value [$I = H(|A_+|)= H(|A_-|)$] as $\gamma
\rightarrow 1^-$. 

\begin{figure}[H]
\begin{center}
\psfig{figure=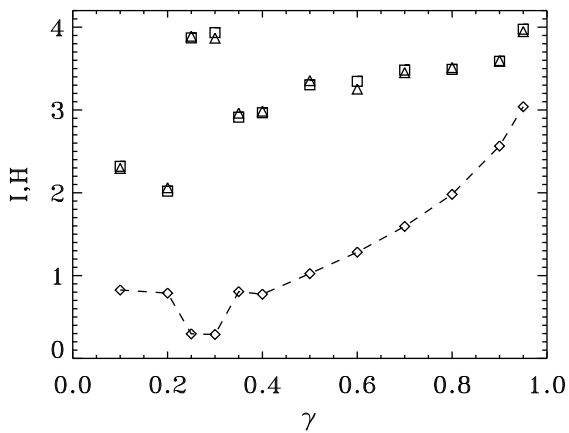,width=.5\textwidth}
\caption[6]{Entropy of $|A_+|$ (squares) and $|A_-|$ (triangles) and their 
mutual
information $I$ (diamonds) as functions of the coupling parameter $\gamma$ 
for $\alpha$ and 
$\beta$ as in Fig \ref{2d}.}
\label{mutua}
\end{center}
\end{figure}

\section{Conclusions}
We have described generalized synchronization phenomena in the 
spatiotemporal dynamics of 
two complex fields which are independent components of a vector complex field. 
Synchronized dynamics of the amplitudes occurs via localized objects. 
The motion of these structures produce the dynamical disorder. 
In $d=1$ localized structures are always dynamical and have the form of 
anticorrelated pulse-hole 
structures. Hole-hole type structures are seen to decay after a short 
transient. In $d=2$ 
the hole-hole type structures become topological vectorial defects which are 
stable for 
small coupling and produce frozen configurations. For coupling above a 
threshold value these
vectorial defects disappear and the persistent dynamics is governed by 
structures which are topological defects of only one of the fields. These 
$d=2$ structures show strong amplitude-amplitude 
anticorrelation and they
are the analog of the $d=1$ pulse-hole structures.
A quantitative characterization of the degree of synchronization is given by 
mutual information 
measures.

\section{Acknowledgments} Financial support from DGYCIT Project PB94-1167 (Spain) is acknowledged.
R.M. also acknowledges financial support from
CONICYT-Fondo Clemente Estable (Uruguay).  M.H. also acknowledges financial support
from the FOMEC project 290, Dep. de F\'{\i}sica FCEyN, Universidad Nacional de
Mar del Plata (Argentina).


}
\end{twocolumns}


\begin{thebibliography}{}


\bibitem[\protect\citeauthoryear{Amengual \bgroup \em et al.\egroup}{1997}]
{prl} Amengual, A., Hern\'andez-Garc\'{\i}a, E., Montagne, R. and San Miguel,
M.  [1997] ``Synchronization of Spatiotemporal Chaos: The Regime of Coupled
Spatiotemporal  Intermittency,'' {\em Phys. Rev. Lett.} {\bf 78}, 4379-4382. 

\bibitem[\protect\citeauthoryear{Amengual \bgroup \em et al.\egroup}{1996}]
{ToniDanielPRL} Amengual, A., Walgraef, D., San Miguel, M., and
Hern\'andez-Garc\'{\i}a [1996] ``Wave Unlocking Transition in Resonantly
Coupled Complex Ginzburg-Landau Equations,'' {\em Phys. Rev. Lett.} {\bf 76},
1956-1959. 

\bibitem[\protect\citeauthoryear{Chat\'e \& Manneville}{1996}]
{Chate96} Chat\'e, H. \& Manneville, P. [1996] ``Phase diagram of the
two-dimensional complex Ginzburg-Landau equation,'' {\em Physica} 
{\bf A 224}, 348-368.

\bibitem[\protect\citeauthoryear{Chat\'e}{1994a}]{Chate94n}
Chat\'e, H. [1994a] ``Spatiotemporal intermittency regimes of the one-dimensional
complex Ginzburg-Landau equation,'' {\em Nonlinearity} {\bf 7}, 185-204.

\bibitem[\protect\citeauthoryear{Chat\'e}{1994b}]{Chate94d}
Chat\'e, H. [1994b]  ``Disordered regimes of the one-dimensional complex
Ginzburg-Landau equation,'' in {\em Spatio-Temporal Patterns in
Non-equilibrium Complex Systems}, eds P.E. Cladis \& P.
Palffy-Muhoray (Addison-Wesley, Reading, MA).

\bibitem[\protect\citeauthoryear{Eguiluz \bgroup \em et al.\egroup}{1998}]
{Eguiluz} Egu\'\i luz, V.M., Hern\'andez-Garc\'\i a, E., \& Piro, O. [1998]
``Boundary effects in the complex Ginzburg-Landau equation,'' This issue.

\bibitem[\protect\citeauthoryear{Gil}{1993}]{Gil} L. Gil [1993] ``Vector Order
Parameter for an Unpolarized Laser and its Vectorial Topological Defects,''
{\em Phys. Rev. Lett.} {\bf 70}, 162-165. 

\bibitem[\protect\citeauthoryear{Hagan}{1982}]{Hagan} Hagan, P.S. [1982]
``Spiral waves in reaction-diffusion equations,'' {\em SIAM J. Appl. Math} 
{\bf
42}, 762-786.

\bibitem[\protect\citeauthoryear{Huygens}{1665}]{huygens}
Huygens, C. [1665] {\em J. Scavants} {\bf XI}, 79-80; {\bf XII}, 86.

\bibitem[\protect\citeauthoryear{Kocarev and Parlitz}{1996a}]{kocarev}
Kocarev, L. and Parlitz, U. [1996] ``Generalized synchronization,
predictability, and equivalence of unidirectionally coupled dynamical 
systems,'' {\em Phys. Rev. Lett.}, {\bf 76},
1816-1819.

\bibitem[\protect\citeauthoryear{Kocarev and Parlitz}{1996b}]{kocarevstc}
Kocarev, L. and Parlitz, U. [1996] ``Synchronizing spatiotemporal chaos in
coupled nonlinear oscillators,'' {\em Phys. Rev. Lett.}, {\bf 77},
2206-2209.

\bibitem[\protect\citeauthoryear{Pecora and Carroll}{1990}]{pecora90}
Pecora, L.M. and Carroll, T.L. [1990], ``Synchronization in chaotic systems,'' 
{\em Phys. Rev. Lett.} {\bf 64}, 821.

\bibitem[\protect\citeauthoryear{Pismen}{1992}]{Pismen92}  Pismen, L. [1992] 
``On
interaction of spiral waves,'' {\em Physica} {\bf D 54}, 183-193.

\bibitem[\protect\citeauthoryear{Pismen}{1994}]{Pismen94} 
Pismen, L. [1994] ``Energy versus Topology: Competing Defect Structures in 2D
Complex Vector Field,'' {\em Phys. Rev. Lett.} {\bf 72}, 2557-2560; 

\bibitem[\protect\citeauthoryear{Rosenblum \bgroup \em et al.\egroup}{1996}] 
{rosenblum} Rosenblum, M.G., Pikovsky, A.S. and Kurths, J. [1996] ``Phase
synchronization of chaotic oscillators,'' {\em Phys. Rev. Lett.} {\bf 76},
1804-1807.

\bibitem[\protect\citeauthoryear{Rulkov \bgroup \em et
al.\egroup}{1995}]{rulkov} Rulkov, N.F., Sushchik, M.M., Tsimring, T.S. and
Abarbanel, H.D.I. [1995] ``Generalized synchronization of chaos in directionally
coupled chaotic systems,'' {\em Phys. Rev. E} {\bf 51}, 980-994.

\bibitem[\protect\citeauthoryear{San Miguel}{1996}]{maxiPRL} San Miguel, M.
[1996] ``Phase Instabilities in the Laser Vector Complex Ginzburg Landau
Equation,'' {\em Phys. Rev. Lett.} {\bf 75}, 425-428.

\bibitem[\protect\citeauthoryear{Shraiman \bgroup \em et al.\egroup}{1992}]
{Shraiman}
Shraiman, B.I., Pumir, A., van Saarlos, W., Hohenberg, P.C, Chat\'e, H.
\& Holen, M. [1992] ``Spatiotemporal chaos in the one-dimensional
Ginzburg-Landau equation,'' {\em Physica} {\bf D 57}, 241-248.

\bibitem[\protect\citeauthoryear{Strogatz and Steward}{1993}]{strogatz}
Strogatz, S.H. and Steward, I., [1993] ``Coupled Oscillators and Biological 
Synchronization,'' {\em Scientific American} {\bf 269}, No. 6, 102.

\bibitem[\protect\citeauthoryear{van Hecke \bgroup \em et al.\egroup}{1998}]
{vHvS} van Hecke, M., Storm, C., van Saarloos, W. [1998] ``Sources, sinks and
wavenumber selection in coupled CGL equations and experimental implications for
counter-propagating wave systems,'' preprint. 

\bibitem[\protect\citeauthoryear{van Hecke}{1998}]{vH}
van Hecke, M. [1998] ``The building blocks of spatiotemporal intermittency,'' 
{\em Phys. Rev. Lett.} {\bf 80}, 1896-1899. 

\bibitem[\protect\citeauthoryear{van Saarloos}{1994}]{vSaarloos}
van Saarloos, W. [1994] ``The Complex Ginzburg-Landau Equation for Beginners,'' 
in {\em Spatio-Temporal Patterns in Non-equilibrium Complex Systems}, eds P.E. 
Cladis \& P. Palffy-Muhoray (Addison-Wesley, Reading, MA).

\bibitem[\protect\citeauthoryear{Winfree}{1980}]{winfree}
Winfree, A.T. [1980] {\em The Geometry of Biological Time}, Springer (New York).
\end{thebibliography}
\end{document}